# Nonclassical Light in a Three-Waveguide Coupler with Second-Order Nonlinearity

Mohd Syafiq M. Hanapi[1], Abdel-Baset M. A. Ibrahim[1,*], Rafael Julius[2], Pankaj K. Choudhury[3], and Hichem Eleuch[4,5]

[1]*Faculty of Applied Sciences, Universiti Teknologi MARA (UiTM), 40450 Shah Alam, Selangor, Malaysia*
[2]*Faculty of Applied Sciences, Universiti Teknologi MARA (UiTM) Perak, Tapah Campus, Perak, Malaysia*
[3]*International Research Center or Advanced Photonics, Zhejiang University, Building 1A, 718 East Haizhou Rd., Haining, Zhejiang 314400, P.R. China*
[4]*Department of Applied Physics and Astronomy, University of Sharjah, Sharjah, United Arab Emirates*
[5]*College of Arts and Sciences, Abu Dhabi University, Abu Dhabi 59911, United Arab Emirates*

***Corresponding author:** abdelbaset@uitm.edu.my; abdelbaset.ibrahim@gmail.com

**Abstract.**
Possible squeezed states generated in a three-waveguide nonlinear coupler operating with second harmonic generation is discussed. This study is carried out using two well-known techniques; the phase space method (based on positive P-representation) and the Heisenberg-based analytical perturbative method. The effect of the key design parameters is analyzed for both codirectional and contra-directional propagation. The optimal degree of feasible squeezing is identified. Also, the performance and capacities of both methods are critically evaluated. For low levels of key design parameters and in the early stages of evolution, a high level of agreement between the two methods is noticed. In the new era of quantum-based technology, the proposed system opens a new avenue for utilising nonlinear couplers in nonclassical light generation.

# 1. Introduction

The coherent state has a quantum noise that is minimal and equally distributed among the two quadratures (for example, amplitude and phase) of the electric field. Coherent light has been used in a variety of applications, including medicine(1), communication (2,3), and industry (4). Yet, it is not enough to rely just on coherent light to drive a new era of technological advancement. For instance, the success of the interferometric method in detecting a gravitational wave (a major step forward for gravitational-wave astronomy) requires squeezed light (5). Squeezed light is a Nonclassical light that is typically produced from the coherent state or vacuum state (of light) by specific optical nonlinear interactions, and it displays reduced noise in one of the two electric field quadrature components. Squeezed states of light have such unique noise distribution where at least one field quadrature falls below the shot noise level (6).

The conventional quantum noise limit must be surpassed to achieve such sensitivity in detecting a gravitational wave (7–11) – a unique characteristic of squeezed light, essential to increase accuracy in a wide variety of applications, including phase measurement (12,13), rotating angle measurement (14), magnetometry (15), and clock synchronization (16). Apart from these, squeezed light can also be employed in medical imaging (17,18) and in the improvement of optical communication systems (19).

A nonlinear interaction is needed to create squeezed light. It may be produced by a variety of nonlinear processes utilizing various devices (20–23). Among many others, the use of a nonlinear coupler (NLC) (24–26) remains limited. A coupler is essentially two adjacent waveguides, each carrying one or more optical modes from a laser source, which exchange energy by coupling the evanescent waves. Quantum nonlinear optical couplers are simple and experimentally realizable. These integrate well with quantum optical devices, such as quantum circuits, photonic chips, and all-optical logic gates (27,28). In addition, the waveguide interaction length and coupling in the nonlinear coupler can also be used to regulate quantum phenomena.

We may classify NLC into two distinct classes based on the number of waveguides: first, the typical two-waveguide devices that have been extensively investigated (29–32), and second, the recently suggested multi-waveguide class of devices (33–38). The two-waveguide class of NLCs fundamentally shows nonclassical behaviour. Nonetheless, the recently proposed multichannel NLCs have shown a higher degree of nonclassicality due to rich Key design parameters and quantum interaction between a high number of propagating modes. The source of nonlinearity in NLCs can be the second (39) or third-order (40), or even a combination of both the linear and nonlinear (41). Third-order nonlinear materials can set off a wide range of nonclassical effects. Yet, it is the second-order nonlinear medium that has a greater impact. Therefore, we believe that a multi-waveguide NLC operating with the second-harmonic generation (SHG) should be a good nonclassical light generator.

To investigate the nonclassical features of light traveling through quantum systems, two kinds of approaches are mostly adopted. The Heisenberg picture forms the foundation for the first kind, in which the field operators change over time, but the state vector remains static. Here, we employ the useful analytical perturbative method, which comes under the Heisenberg picture and is regarded as an enhanced form of the common short-length approximation method. Quantum-coupled nonlinear differential equations are then obtained using the Heisenberg equation of motion. The second method is based on the Schrödinger picture where the operators remain fixed, but the state vector changes with time. Here, we also employ the phase space approach, which is a key technique under the Schrodinger picture. By substituting the correct Hamiltonian for the system into the von Neumann equation of motion, we can construct a nonlinear differential equation of the density matrix. The nonclassical states of the system are often exclusively explored with one of these two methods in almost all prior research on quantum NLCs. Nevertheless, rather than depending on a single theoretical prediction, we use both methods to investigate the squeezed states of light propagating in a three-waveguided NLC with SHG. This will not only ensure accurate findings but will provide insights into the strengths and weaknesses of each technique used.

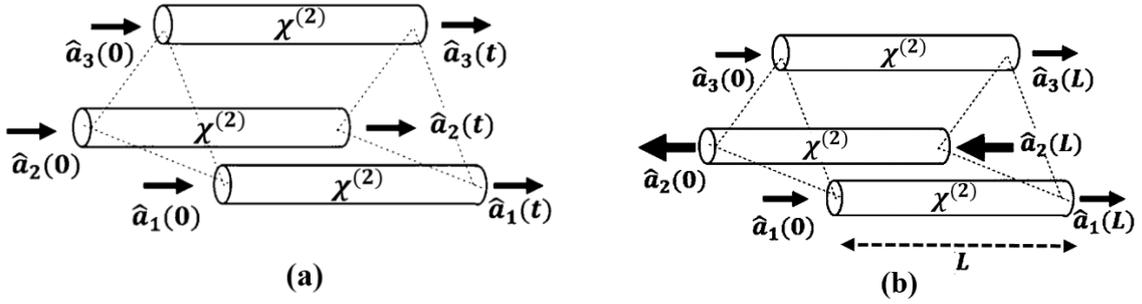

**Figure 1.** Basic diagram of the three-channel nonlinear coupler with second-order nonlinearity (a) codirectional propagation. (b) A contra-directional system where the mode $\hat{a}_2$ propagates opposite to the other two modes

In this study, we employ both the analytical perturbative techniques and the phase space method to examine the single-mode squeezing in three-channel NLC waveguides with second-order nonlinearity. A single fundamental mode travels across each waveguide with fundamental frequency ω. The fundamental photons of the pump field also generate second harmonic (SH) modes propagating along the fiber with double frequency (2ω). Figure 1 depicts a schematic illustration of the system under consideration for the case of codirectional (Fig. 1a) and contra-directional propagation (Fig. 1b). In codirectional propagation, all three modes propagate in the same direction, whereas in contra-directional propagation, one mode (Mode 2) is assumed to propagate in opposite direction. The effect of key design parameters on the generated squeezed states is examined and the optimal degree of possible squeezing is reported. Also, the performance and capacities of both methods are critically evaluated. The current triple-waveguide structure may provide a more efficient mechanism for generating nonclassical effects with enhanced performance. This is because its coupled-mode interactions and correlations are more adaptable.

## 2. Mathematical Formulation

Here, we describe the mathematical formulation of the system under consideration using both methods. We first construct the Hamiltonian which properly describes the system under consideration. For three-channel NLC devices, such Hamiltonian can be constructed as follows:

$$\begin{aligned}\hat{H} &= \hat{H}_{free} + \hat{H}_{Linear} + \hat{H}_{Nonlinear} + \hat{H}_{SHG} \\ &= \hbar\left[\omega_1 \hat{a}_1^\dagger \hat{a}_1 + \omega_2 \hat{a}_2^\dagger \hat{a}_2 + \omega_3 \hat{a}_3^\dagger \hat{a}_3\right] + \hbar\kappa\left[\hat{a}_1^\dagger \hat{a}_2 + \hat{a}_2^\dagger \hat{a}_3 + \hat{a}_3^\dagger \hat{a}_1 + \text{h.c.}\right] \\ &+ \frac{ig}{2}\hbar\left[\left(\hat{a}_1^{\dagger 2}\hat{b}_1 + \hat{a}_2^{\dagger 2}\hat{b}_2 + \hat{a}_3^{\dagger 2}\hat{b}_3 - \text{h.c.}\right)\right] + \hbar\left[(2\omega_1)\hat{b}_1^\dagger \hat{b}_1 + (2\omega_2)\hat{b}_2^\dagger \hat{b}_2 + (2\omega_3)\hat{b}_3^\dagger \hat{b}_3\right]\end{aligned} \qquad (1)$$

The free Hamiltonian $\hat{H}_{free}$ includes three fundamental modes ($\hat{a}_1, \hat{a}_2, \hat{a}_3$) propagating in their channels with fundamental frequencies $\omega_1$, $\omega_2$, and $\omega_3$, respectively. The Hamiltonian linear coupling $\hat{H}_{Linear}$ represents the linear coupling between each one of the three modes and the strength of the coupling is quantified by the linear coupling coefficient $\kappa$ while h.c. stands for the Hermitian conjugate. This coupling is due to the overlapping of the evanescent waves of each propagating field mode. The $\hat{H}_{Nonlinear}$ represents the second-order nonlinear interaction between each mode and the relevant waveguide, and the strength of coupling is quantified by the nonlinear coupling coefficient g. Finally, the Hamiltonian $\hat{H}_{SHG}$ determines three SH modes ($\hat{b}_1, \hat{b}_2, \hat{b}_3$) propagating with frequencies $2\omega_1$, $2\omega_2$, and $2\omega_3$, respectively. The second-order nonlinear interaction process generates these modes with frequencies double those of the fundamental modes.

*The Phase-Space Method*

To describe the dynamics of the density matrix evolution, the phase space method substitutes the full Hamiltonian into the von Neumann equation of motion. Following that, one of the representations might be used to transform the master equation to its classical Fokker-Planck (FP) equation in phase space. In this work, we use the positive P-representation. An initial time-dependent distribution function P is guaranteed to exist, to be positive, and to fulfill an FP equation in the positive-P representation. One method for solving the FP equation is to use Ito rules to determine the equivalent set of noisy coupled stochastic equations. The phase space variables utilized to compute the required nonclassical states can be obtained by numerically solving these stochastic equations. We recall the standard form of the Liouville-Von Neumann equation.

$$i\hbar \frac{\partial \hat{\rho}}{\partial t} = \left[\hat{H}, \hat{\rho}\right] \qquad (2)$$

Substituting the total Hamiltonian from Eq. (1) into Eq. (2) yields the reduced dynamical equation of the density operator as

$$\frac{\partial \hat{\rho}}{\partial t} = i\omega_1 \left( \hat{\rho}\hat{a}_1^\dagger \hat{a}_1 - \hat{a}_1^\dagger \hat{a}_1 \hat{\rho} \right) + i\omega_2 \left( \hat{\rho}\hat{a}_2^\dagger \hat{a}_2 - \hat{a}_2^\dagger \hat{a}_2 \hat{\rho} \right) + i\omega_3 \left( \hat{\rho}\hat{a}_3^\dagger \hat{a}_3 - \hat{a}_3^\dagger \hat{a}_3 \hat{\rho} \right)$$
$$+ i\kappa \left( \hat{\rho}\hat{a}_1^\dagger \hat{a}_2 - \hat{a}_1^\dagger \hat{a}_2 \hat{\rho} \right) + i\kappa \left( \hat{\rho}\hat{a}_1 \hat{a}_2^\dagger - \hat{a}_1 \hat{a}_2^\dagger \hat{\rho} \right) + i\kappa \left( \hat{\rho}\hat{a}_2^\dagger \hat{a}_3 - \hat{a}_2^\dagger \hat{a}_3 \hat{\rho} \right)$$
$$+ i\kappa \left( \hat{\rho}\hat{a}_2 \hat{a}_3^\dagger - \hat{a}_2 \hat{a}_3^\dagger \hat{\rho} \right) + i\kappa \left( \hat{\rho}\hat{a}_3^\dagger \hat{a}_1 - \hat{a}_3^\dagger \hat{a}_1 \hat{\rho} \right) + i\kappa \left( \hat{\rho}\hat{a}_3 \hat{a}_1^\dagger - \hat{a}_3 \hat{a}_1^\dagger \hat{\rho} \right)$$
$$+ \frac{g}{2} \left( \hat{a}_1^{\dagger 2} \hat{b}_1 \hat{\rho} - \hat{\rho}\hat{a}_1^{\dagger 2} \hat{b}_1 \right) + \frac{g}{2} \left( \hat{a}_2^{\dagger 2} \hat{b}_2 \hat{\rho} - \hat{\rho}\hat{a}_2^{\dagger 2} \hat{b}_2 \right) + \frac{g}{2} \left( \hat{a}_3^{\dagger 2} \hat{b}_3 \hat{\rho} - \hat{\rho}\hat{a}_3^{\dagger 2} \hat{b}_3 \right) \quad (3)$$
$$+ \frac{g}{2} \left( \hat{\rho}\hat{a}_1^2 \hat{b}_1^\dagger - \hat{a}_1^2 \hat{b}_1^\dagger \hat{\rho} \right) + \frac{g}{2} \left( \hat{\rho}\hat{a}_2^2 \hat{b}_2^\dagger - \hat{a}_2^2 \hat{b}_2^\dagger \hat{\rho} \right) + \frac{g}{2} \left( \hat{\rho}\hat{a}_3^2 \hat{b}_3^\dagger - \hat{a}_3^2 \hat{b}_3^\dagger \hat{\rho} \right)$$
$$+ i2\omega_1 \left( \hat{\rho}\hat{b}_1^\dagger \hat{b}_1 - \hat{b}_1^\dagger \hat{b}_1 \hat{\rho} \right) + i2\omega_2 \left( \hat{\rho}\hat{b}_2^\dagger \hat{b}_2 - \hat{b}_2^\dagger \hat{b}_2 \hat{\rho} \right) + i2\omega_3 \left( \hat{\rho}\hat{b}_3^\dagger \hat{b}_3 - \hat{b}_3^\dagger \hat{b}_3 \hat{\rho} \right)$$

Equation (3) is a quantum mechanical partial differential equation. It describes the time evolution of the density operator $\hat{\rho}$. However, this equation is difficult to solve. The standard method in quantum optics is to convert it to the corresponding classical c-number FP equation using one of the available representations such as positive-P or Wigner. Here we employ the positive-P representation with these quantum-classical operator correspondences (42,43).

$$\hat{a}^\dagger \hat{\rho} = \left( \beta - \frac{\partial}{\partial \alpha} \right) P, \hat{a}\hat{\rho} = \alpha P, \hat{\rho}\hat{a} = \left( \alpha - \frac{\partial}{\partial \beta} \right) P, \hat{\rho}\hat{a}^\dagger = \beta P \quad (4)$$

The quantum-classical correspondences in positive-P representation map the evolution of the density matrix $\hat{\rho}$ to a classical probability distribution $P(\alpha, \beta, t)$ in phase space where $\alpha$ and $\beta$ are independent complex variables. The FP equation may then be expressed as a collection of noisy stochastic equations obeying Ito principles, which is a standard approach in statistical mechanics (44). For the present system, the following stochastic differential equations result from the process:

$$\frac{d\alpha_1}{dt} = -i\omega_1 \alpha_1 + g\bar{\alpha}_1 \beta_1 - i\kappa\alpha_2 - i\kappa\alpha_3 + \sqrt{g\bar{\alpha}_1}\eta_1(t) \quad (5)$$

$$\frac{d\beta_1}{dt} = i\omega_1 \beta_1 + g\alpha_1 \bar{\beta}_1 + i\kappa\beta_2 + i\kappa\beta_3 + \sqrt{g\bar{\beta}_1}\eta_2(t) \quad (6)$$

$$\frac{d\alpha_2}{dt} = -i\omega_2 \alpha_2 + g\bar{\alpha}_2 \beta_2 - i\kappa\alpha_1 - i\kappa\alpha_3 + \sqrt{g\bar{\alpha}_2}\eta_3(t) \quad (7)$$

$$\frac{d\beta_2}{dt} = i\omega_2 \beta_2 + g\alpha_2 \bar{\beta}_2 + i\kappa\beta_1 + i\kappa\beta_3 + \sqrt{g\bar{\beta}_2}\eta_4(t) \quad (8)$$

$$\frac{d\alpha_3}{dt} = -i\omega_3 \alpha_3 + g\bar{\alpha}_3 \beta_3 - i\kappa\alpha_2 - i\kappa\alpha_1 + \sqrt{g\bar{\alpha}_3}\eta_5(t) \quad (9)$$

$$\frac{d\beta_3}{dt} = i\omega_3 \beta_3 + i\kappa\beta_2 + g\alpha_3 \bar{\beta}_3 + i\kappa\beta_1 + \sqrt{g\bar{\beta}_3}\eta_6(t) \quad (10)$$

$$\frac{d\bar{\alpha}_1}{dt} = -i(2\omega_1)\bar{\alpha}_1 - \frac{g}{2}\alpha_1^2 \quad (11)$$

$$\frac{d\bar{\beta}_1}{dt} = i(2\omega_1)\bar{\beta}_1 - \frac{g}{2}\beta_1^2 \quad (12)$$

$$\frac{d\bar{\alpha}_2}{dt} = -i(2\omega_2)\bar{\alpha}_2 - \frac{g}{2}\alpha_2^2 \tag{13}$$

$$\frac{d\bar{\beta}_2}{dt} = i(2\omega_2)\bar{\beta}_2 - \frac{g}{2}\beta_2^2 \tag{14}$$

$$\frac{d\bar{\alpha}_3}{dt} = -i(2\omega_3)\bar{\alpha}_3 - \frac{g}{2}\alpha_3^2 \tag{15}$$

$$\frac{d\bar{\beta}_2}{dt} = i(2\omega_3)\bar{\beta}_2 - \frac{g}{2}\beta_3^2 \tag{16}$$

In Eqs. (5)–(16), we can simplify the picture by considering that $\alpha_i$ and $\beta_i$ are the classical equivalence of the operators $\hat{a}_i$ and $\hat{a}_i^\dagger$ ($i$ = 1, 2, and 3), respectively, while $\bar{\alpha}_i$ and $\bar{\beta}_i$ are the classical equivalent of the operators $\hat{b}_i$ and $\hat{b}^\dagger{}_i$. Therefore, for the three fundamental modes, we have three sets of variables ($\alpha_i$, $\beta_i$) while for the three SH modes, we have another three sets of these classical complex variables $(\bar{\alpha}_i, \bar{\beta}_i)$. Note that, Eqs. (5)–(10) contain fluctuating forces $\eta_i(t)$ ($i$ = 1, 2, 3, 4, 5 and 6). These fluctuations have zero mean and are correlated in time $\langle \eta_i(t)\eta_j(t')\rangle = \delta_{ij}\delta(t-t')$. As a result, the noise will dominate any single trajectory solution to these equations. A stable solution can only be obtained by averaging it across many trajectories.

*The Analytical Perturbative Method*

The analytical perturbative method has been used to study the quantum behaviour of a variety of systems (45–47). This technique has time-efficient numerical computation since it requires a single operation only. In this approach, Quantum coupled nonlinear differential equations are obtained by plugging in the proper Hamiltonian or the momentum operator into the Heisenberg equation of motion. The solution of these coupled equations of motion is assumed in the form of the Baker–Hausdorff (BH) formula, which is then expanded as a Taylor series up to the second order, and the intuitive analytical solution is proposed based on that expansion. As this technique is Schrödinger-picture based, the evolution of each mode is individually described through the quantum mechanical Heisenberg equation of motion, generally expressed as

$$i\hbar \frac{d\hat{a}_j}{dt} = \left[\hat{a}_j, \hat{H}\right] \tag{17}$$

Substituting the Hamiltonian from Eq. (1) into Eq. (17) for modes $\hat{a}_1, \hat{a}_2, \hat{a}_3, \hat{b}_1, \hat{b}_2$ and $\hat{b}_3$, we obtain the following coupled system of equations, which describes the time evolution the modes, as follows:

$$\frac{d\hat{a}_1}{dt} = i\omega_1\hat{a}_1 + g\hat{a}_1^\dagger\hat{b}_1 + i\kappa\hat{a}_2 + i\kappa\hat{a}_3 \tag{18}$$

$$\frac{d\hat{a}_2}{dt} = i\omega_2\hat{a}_2 + g\hat{a}_2^\dagger\hat{b}_2 + i\kappa\hat{a}_1 + i\kappa\hat{a}_3 \tag{19}$$

$$\frac{d\hat{a}_3}{dt} = i\omega_3 \hat{a}_3 + g\hat{a}_3^\dagger \hat{b}_3 + i\kappa \hat{a}_2 + i\kappa \hat{a}_1 \tag{20}$$

$$\frac{d\hat{b}_1}{dt} = i(2\omega_1)\hat{b}_1 - \frac{g}{2}\hat{a}_1^2 \tag{21}$$

$$\frac{d\hat{b}_2}{dt} = i(2\omega_2)\hat{b}_2 - \frac{g}{2}\hat{a}_2^2 \tag{22}$$

$$\frac{d\hat{b}_3}{dt} = i(2\omega_3)\hat{b}_3 - \frac{g}{2}\hat{a}_3^2 \tag{23}$$

Eqs. (18)–(20) in the previous set of coupled equations describe the propagation of fundamental modes while Eqs. (21)–(23) describe the propagation of SH modes. Solutions to these equations are assumed to satisfy the Baker-Campbell-Hausdorff (BCH) formula, as follows.

$$\hat{a}_j = \exp\left(\frac{i}{\hbar}\hat{H}t\right)\hat{a}_j(0)\exp\left(-\frac{i}{\hbar}\hat{H}t\right) = \hat{a}_j(0) + \frac{it}{\hbar}\left[\hat{H},\hat{a}_j(0)\right] - \frac{1}{2}\frac{t^2}{\hbar^2}\left[\hat{H},\left[\hat{H},\hat{a}_j(0)\right]\right] + \ldots \tag{24}$$

For each mode ($\hat{a}_1$, $\hat{a}_2$, $\hat{a}_3$, $\hat{b}_1$, $\hat{b}_2$ and $\hat{b}_3$), the commutation relations $\left[\hat{H},\hat{a}_j(0)\right]$ and $\left[\hat{H},\left[\hat{H},\hat{a}_j(0)\right]\right]$ in Eq. (24) is evaluated utilizing Eqs. (18)–(23) to obtain the intuitive mode solutions of the following forms:

$$\begin{aligned}\hat{a}_1(t) = &\hat{a}_1(0)A_1 + \hat{a}_2(0)A_2 + \hat{a}_3(0)A_3 + \hat{a}_1^\dagger(0)\hat{b}_1(0)A_4 + \hat{a}_2^\dagger(0)\hat{b}_2(0)A_5 \\ &+\hat{a}_3^\dagger(0)\hat{b}_3(0)A_6 + \hat{a}_2^\dagger(0)\hat{b}_1(0)A_7 + \hat{a}_3^\dagger(0)\hat{b}_1(0)A_8 + A_9\,\hat{a}_1\hat{b}_1^\dagger\hat{b}_1 + A_{10}\,\hat{a}_1^\dagger\hat{a}_1^2\end{aligned} \tag{25}$$

$$\begin{aligned}\hat{a}_2(t) = &\hat{a}_1(0)B_1 + \hat{a}_2(0)B_2 + \hat{a}_3(0)B_3 + \hat{a}_1^\dagger(0)\hat{b}_2(0)B_4 + \hat{a}_1^\dagger(0)\hat{b}_1(0)B_5 \\ &+\hat{a}_2^\dagger(0)\hat{b}_2(0)B_6 + \hat{a}_3^\dagger(0)\hat{b}_3(0)B_7 + \hat{a}_3^\dagger(0)\hat{b}_2(0)B_8 + B_9\,\hat{a}_2\hat{b}_2^\dagger\hat{b}_2 + B_{10}\,\hat{a}_2^\dagger\hat{a}_2^2\end{aligned} \tag{26}$$

$$\begin{aligned}\hat{a}_3(t) = &\hat{a}_1(0)C_1 + \hat{a}_2(0)C_2 + \hat{a}_3(0)C_3 + \hat{a}_1^\dagger(0)\hat{b}_3(0)C_4 + \hat{a}_2^\dagger(0)\hat{b}_2(0)C_5 \\ &+\hat{a}_1^\dagger(0)\hat{b}_1(0)C_6 + \hat{a}_3^\dagger(0)\hat{b}_3(0)C_7 + \hat{a}_3(0)\hat{b}_3^\dagger(0)\hat{b}_3(0)C_8 + C_9\,\hat{a}_2^\dagger\hat{b}_3 + C_{10}\,\hat{a}_3^\dagger\hat{a}_3^2\end{aligned} \tag{27}$$

$$\begin{aligned}\hat{b}_1(t) = &D_1\hat{b}_1(0) + D_2\,\hat{a}_1^2(0) + D_3\,\hat{a}_2(0)\hat{a}_1(0) + D_4\,\hat{a}_3(0)\hat{a}_1(0) \\ &+ D_5\,\hat{a}_1^\dagger(0)\hat{a}_1(0)\hat{b}_1(0) + D_6\,\hat{a}_1(0)\hat{a}_1^\dagger(0)\hat{b}_1(0)\end{aligned} \tag{28}$$

$$\begin{aligned}\hat{b}_2(t) = &E_1\hat{b}_2(0) + E_2\,\hat{a}_2^2(0) + E_3\,\hat{a}_1(0)\hat{a}_2(0) + E_4\,\hat{a}_3(0)\hat{a}_2(0) \\ &+ E_5\,\hat{a}_2^\dagger(0)\hat{a}_2(0)\hat{b}_2(0) + E_6\,\hat{a}_2(0)\hat{a}_2^\dagger(0)\hat{b}_2(0)\end{aligned} \tag{29}$$

$$\begin{aligned}\hat{b}_3(t) = &F_1\hat{b}_3(0) + F_2\,\hat{a}_2(0)\hat{a}_3(0) + F_3\,\hat{a}_1(0)\hat{a}_3(0) + F_4\,\hat{a}_3^2(0) \\ &+ F_5\,\hat{a}_3^\dagger(0)\hat{a}_3(0)\hat{b}_3(0) + F_6\,\hat{a}_3(0)\hat{a}_3^\dagger(0)\hat{b}_3(0)\end{aligned} \tag{30}$$

In the analytical method, it is common to assume a weak nonlinear interaction in the form of a perturbation and ignore the higher-order nonlinear coupling terms. However, in the previous mode solutions in Eqs. (25)–(30), we have retained all nonlinear terms, i.e., both terms containing nonlinear coefficients $g$ and those with $g^2$. The coefficients $\{A_k(t)\}_1^{10}$, $\{B_k(t)\}_1^{10}$, $\{C_k(t)\}_1^{10}$, $\{D_k(t)\}_1^{6}$, $\{E_k(t)\}_1^{6}$ and $\{F_k(t)\}_1^{6}$ are all time-dependent. In the well-known short-length approximation method, these coefficients are essentially approximated by the second-degree polynomials in time. The analytical perturbative method is superior to the short-length approximation method because these coefficients are evaluated precisely using additional mathematical steps. Basically, in the perturbative approach, sets of coupled differential equations describing these coefficients are derived by substituting the previous mode solutions in Eqs. (25)– (30) back into the mode evolution Eqs. (18)– (23) and equating similar terms on both sides. To obtain the coefficients, these sets of equations are numerically solved (please see Appendix A).

*Calculation of Single-Mode Squeezing*

Squeezed states are those in which the uncertainty distribution in one quadrature component is reduced below the standard noise limit of a coherent state at the expense of increased uncertainty in the other quadrature, thereby ensuring that the Heisenberg uncertainty relation is not violated (48). Mathematically, squeezing is indicated by one component of quadrature variances that has a value below the shot noise level (49). The operators $\hat{X}_j = \frac{1}{2}(\hat{a}_1 + \hat{a}_1^\dagger)$ and $\hat{Y}_j = \frac{1}{2i}(\hat{a}_1 - \hat{a}_1^\dagger)$ are the optical field quadrature operators and $\langle(\Delta\hat{X}_j)^2\rangle$, $\langle(\Delta\hat{Y}_j)^2\rangle$ are their variances, squeezing is identified by having one of the quadrature components a value less than the shot noise level., i.e., $\langle(\Delta\hat{X}_j)^2\rangle < \frac{1}{4}$ or $\langle(\Delta\hat{Y}_j)^2\rangle < \frac{1}{4}$ where $\langle(\Delta\hat{X}_j)^2\rangle = \langle\hat{X}_j^2\rangle - \langle\hat{X}_j\rangle^2$ and $\langle(\Delta\hat{Y}_j)^2\rangle = \langle\hat{Y}_j^2\rangle - \langle\hat{Y}_j\rangle^2$. These quadrature variances are expressed in terms of the creation and annihilation operators as

$$\begin{bmatrix} \langle(\Delta\hat{X}_j)^2\rangle \\ \langle(\Delta\hat{Y}_j)^2\rangle \end{bmatrix} = \frac{1}{4}\left\{1 + 2\langle\hat{a}_j^\dagger\hat{a}_j\rangle - 2\langle\hat{a}_j^\dagger\rangle\langle\hat{a}_j\rangle \pm \left[\langle\hat{a}_j^2\rangle - \langle\hat{a}_j\rangle^2 + \langle\hat{a}_j^{\dagger 2}\rangle - \langle\hat{a}_j^\dagger\rangle^2\right]\right\} \quad j=1,2,3. \quad (31)$$

In the preceding equation, $j$ is the mode number and the bracket < > represents the normal-ordered expectation value, which necessitates the creation operators to be always to the left of the annihilation operators in any of their products. In the phase space method, the operators $\hat{a}_j$ and $\hat{a}_j^\dagger$ in Eq. (31) are replaced with their classical equivalence, i.e., the complex phase space variables $\alpha_j$ and $\beta_j$ respectively. This yields the following equation of the quadrature variances of the field in phase space:

$$\begin{bmatrix} \langle(\Delta\hat{X}_j)^2\rangle \\ \langle(\Delta\hat{Y}_j)^2\rangle \end{bmatrix} = \frac{1}{4}\left\{1 + 2\langle\beta_j\alpha_j\rangle - 2\langle\beta_j\rangle\langle\alpha_j\rangle \pm \left[\langle\alpha_j^2\rangle - \langle\alpha_j\rangle^2 + \langle\beta_j^2\rangle - \langle\beta_j\rangle^2\right]\right\} \quad j=1,2,3. \quad (32)$$

For the analytical method, expressions for the quadrature variances of the fundamental modes are obtained by substituting the mode solutions from Eqs. (25)–(27) into Eq. (31) and utilizing the classical equivalence $\langle \hat{a}_j(0) \rangle = \alpha_j$ and $\langle \hat{b}_j(0) \rangle = \beta_j$ results in the expressions shown below:

$$\begin{bmatrix} \langle (\Delta \hat{X}_1)^2 \rangle \\ \langle (\Delta \hat{Y}_1)^2 \rangle \end{bmatrix} = \tfrac{1}{4}\Big[ 1 + 2\big(|A_5|^2 |\beta_2|^2 + |A_6|^2 \beta_3^* \beta_3 + A_7 A_5^* \beta_2^* \beta_1 \\ + A_8 A_6^* \beta_3^* \beta_1 + A_5 A_7^* \beta_1^* \beta_2 + A_6 A_8^* \beta_1^* \beta_3 \big) \\ \pm \big( A_4 A_1 \beta_1 + A_{10} A_1 \alpha_1^2 + A_5 A_2 \beta_2 + A_7 A_2 \beta_1 + A_6 A_3 \beta_3 + A_8 A_3 \beta_1 + c.c. \big) \Big] \tag{33}$$

$$\begin{bmatrix} \langle (\Delta \hat{X}_2)^2 \rangle \\ \langle (\Delta \hat{Y}_2)^2 \rangle \end{bmatrix} = \tfrac{1}{4}\Big[ 1 + 2\big(|B_5|^2 |\beta_1|^2 + |B_7|^2 |\beta_3|^2 + B_5 B_4^* \beta_2^* \beta_1 \\ + B_4 B_5^* \beta_1^* \beta_2 + B_8 B_7^* \beta_3^* \beta_2 + B_7 B_8^* \beta_2^* \beta_3 \big) \\ \pm \big( B_4 B_1 \beta_2 + B_5 B_1 \beta_1 + B_6 B_2 \beta_2 + B_{10} B_2 \alpha_2^2 + B_7 B_3 \beta_3 + B_8 B_3 \beta_2 + c.c. \big) \Big] \tag{34}$$

$$\begin{bmatrix} \langle (\Delta \hat{X}_3)^2 \rangle \\ \langle (\Delta \hat{Y}_3)^2 \rangle \end{bmatrix} = \tfrac{1}{4}\Big[ 1 + 2\big(|C_5|^2 |\beta_2|^2 + |C_6|^2 |\beta_1|^2 + C_6 C_4^* \beta_3^* \beta_1 + C_4 C_6^* \beta_1^* \beta_3 + C_5 C_9^* \beta_3^* \beta_2 \big) \\ \pm \big( C_4 C_1 \beta_3 + C_6 C_1 \beta_1 + C_5 C_2 \beta_2 + C_9 C_2 \beta_3 + C_7 C_3 \beta_3 + C_{10} C_3 \alpha_3^2 + c.c. \big) \Big] \tag{35}$$

where c.c. symbolizes the complex conjugate.

*The contra-directional Propagation*

In the investigation of the contra-directional propagation, we follow the same procedure proposed by Perina (50,51) and subsequently by others (52,53). First, the temporal system is converted into a spatial system. The conversion can be done using the time-displacement formula $dt = \frac{dz}{v_c}$, where $v_c$ is the speed mode. Second, an opposite sign is introduced to the spatial equations of the contra-propagating mode $\hat{a}_2$ as shown in Figure 1(b). In phase space method, the conversion is made to the stochastics differential equations (5)-(16) by dividing both sides of each equation by $v_c$ and introducing a negative sign to the spatial equations of the second mode $\hat{a}_2$. In this case, the Stochastic set of differential equations representing the contra-directional case can be written as

$$\frac{d\alpha_1}{dz} = -ik_1\alpha_1 + G\bar{\alpha}_1\beta_1 - iK\alpha_2 - iK\alpha_3 + \sqrt{G\bar{\alpha}_1}\xi_1(z) \tag{36}$$

$$\frac{d\beta_1}{dz} = ik_1\beta_1 + G\alpha_1\bar{\beta}_1 + iK\beta_2 + iK\beta_3 + \sqrt{G\bar{\beta}_1}\xi_2(t) \tag{37}$$

$$\frac{d\alpha_2}{dz} = +ik_2\alpha_2 - G\bar{\alpha}_2\beta_2 + iK\alpha_1 + iK\alpha_3 - \sqrt{G\bar{\alpha}_2}\xi_3(t) \tag{38}$$

$$\frac{d\beta_2}{dz} = -ik_2\beta_2 - G\alpha_2\bar{\beta}_2 - iK\beta_1 - iK\beta_3 - \sqrt{G\bar{\beta}_2}\xi_4(t) \tag{39}$$

$$\frac{d\alpha_3}{dz} = -ik_3\alpha_3 + G\bar{\alpha}_3\beta_3 - iK\alpha_2 - iK\alpha_1 + \sqrt{G\bar{\alpha}_3}\xi_5(t) \tag{40}$$

$$\frac{d\bar{\beta}_3}{dz} = ik_3\bar{\beta}_3 + iK\bar{\beta}_2 + G\alpha_3\bar{\beta}_3 + iK\bar{\beta}_1 + \sqrt{G\bar{\beta}_3}\xi_6(t) \tag{41}$$

$$\frac{d\bar{\alpha}_1}{dz} = -i(2k_1)\bar{\alpha}_1 - \frac{G}{2}\alpha_1^2 \tag{42}$$

$$\frac{d\bar{\beta}_1}{dz} = i(2k_1)\bar{\beta}_1 - \frac{G}{2}\beta_1^2 \tag{43}$$

$$\frac{d\bar{\alpha}_2}{dz} = -i(2k_2)\bar{\alpha}_2 - \frac{G}{2}\alpha_2^2 \tag{44}$$

$$\frac{d\bar{\beta}_2}{dz} = i(2k_2)\bar{\beta}_2 - \frac{G}{2}\beta_2^2 \tag{45}$$

$$\frac{d\bar{\alpha}_3}{dz} = -i(2k_3)\bar{\alpha}_3 - \frac{G}{2}\alpha_3^2 \tag{46}$$

$$\frac{d\bar{\beta}_2}{dz} = i(2k_3)\bar{\beta}_2 - \frac{G}{2}\beta_3^2 \tag{47}$$

In the previous system (36)-(47), we have used $k_j = \frac{\omega_j}{v_c}$, with $j = 1,2,3$, $G = \frac{g}{v_c}$, $K = \frac{\kappa}{v_c}$ and $\xi_l(z) = \frac{\eta_l(t)}{\sqrt{v_c}}$ with $l = 1, 2,...,6$. For the analytical perturbative method, the same process is also performed on the set of equations (A1)-(A48). It is important to note that the Hamiltonian (1) describes the fields at all points of the interaction volume. This means that the spatial description is accurate only when the forward wave reaches $z = L$ and the backward wave reaches $z = 0$. However, it does not accurately describe the transient states of the contra-propagating fields for $0 < z < L$. For all operators in Eq. (1), this formulation guarantees the conservation of the boson commutation rules in both forward and backward propagating fields.

*Numerical Procedure*

To summarize the numerical strategy for the phase space method, the system of Eqs. (5)–(16) is integrated numerically using the fourth order Runge-Kutta method. As explained earlier, averaging across many trajectories is required to obtain a stable solution, which yields numerical values of ($\alpha_i$, $\beta_i$) for the three fundamental modes and $(\bar{\alpha}_i, \bar{\beta}_i)$ for the three SH modes, where $i = 1, 2,$ and 3. These values are used to evaluate the optical field quadrature variances for the propagating modes $\langle(\Delta\hat{X}_j)^2\rangle, \langle(\Delta\hat{Y}_j)^2\rangle$ using Eq. (32). For the analytical method, the numerical procedure involves two main steps – viz. (i) solving simultaneously the coupled sets of Eqs. (A1)–(A48) to obtain numerical values of the time-dependent coefficients $\{A_k(t)\}_1^{10}$, $\{B_k(t)\}_1^{10}$, $\{C_k(t)\}_1^{10}$, $\{D_k(t)\}_1^{6}$, $\{E_k(t)\}_1^{6}$ and $\{F_k(t)\}_1^{6}$, and (ii) with the knowledge of these coefficients, the optical field quadrature variances for the propagating modes $\langle(\Delta\hat{X}_j)^2\rangle, \langle(\Delta\hat{Y}_j)^2\rangle$ can be evaluated using Eqs. (33)–(35). It should be emphasized that ($\alpha_i$, $\beta_i$) and $(\bar{\alpha}_i, \bar{\beta}_i)$ are time-dependent in the phase-space method, whereas they are time-independent in the analytical method.

Finally, for the convenience of numerical simulation, dimensionless input parameters are used. For the codirectional system, the relevant systems of Eqs. (5)–(16) and (A1)–(A48) are scaled with respect to the input frequency of the first mode ($\omega_1$), such that $\tilde{\omega}_1 = \frac{\omega_1}{\omega_1} = 1$, $\tilde{\omega}_2 = \frac{\omega_2}{\omega_1}$, $\widetilde{\omega}_3 = \frac{\omega_3}{\omega_1}$, $\tilde{g} = \frac{g}{\omega_1}$, $\tilde{\kappa} = \frac{\kappa}{\omega_1}$, $\tau = \omega_1 t$ and $\tilde{\eta}(\tau) = \frac{\eta(t)}{\sqrt{\omega_1}}$. For The contra-directional case, the spatial system (36)-(47) can be scaled with the wavenumber of the first mode $k_1$ using the dimensionless parameters, $\tilde{k}_1 = \frac{k_1}{k_1} = 1$, $\tilde{k}_2 = \frac{k_2}{k_1}$, $\tilde{k}_3 = \frac{k_3}{k_1}$, $\tilde{G} = \frac{G}{k_1}$, $\tilde{K} = \frac{K}{k_1}$, $\tilde{z} = k_1 z$ and $\tilde{\tilde{\xi}}_l = \frac{\xi_l}{\sqrt{k_1}}$. This will produce a dimensionless form of the spatial system (36)-(47).

## 3. Result and Discussion

We begin this section by looking at the evolution of the field quadrature variances at various initial input field amplitudes. Other input parameters are fixed, with the nonlinear coupling coefficient of $\tilde{g}$ set to 0.01 and the linear coupling $\tilde{\kappa}$ to 0.1. likewise, the frequency of propagating modes is fixed at $\widetilde{\omega}_1 = \widetilde{\omega}_2 = \widetilde{\omega}_3 = 1$. In Fig. 2, We focus on the case that the initial amplitudes of the input fields are equal; this is known as symmetric initialization (where $\alpha_1 = \alpha_2 = \alpha_3 = 1$). Under this combination of input values, both methods yield a continuous oscillation pattern. Squeezing in the first, second, and third modes is found to be identical, i.e., all three modes yield the same degree of squeezing. Therefore, only the squeezed states generated in the first channel are shown in Fig. 2. At the earlier stages of evolution, both methods acquire a high degree of agreement. Whereas, as time passes, the phase space method predicts a lower amplitude for the field quadratures than the analytical perturbative method. Despite this, both systems display signals that are completely in phase. However, the constant increase in field amplitude found by the analytical method over time does not appear reasonable. This indicates that the analytical method cannot replace the positive P phase space method in all combinations of system parameters.

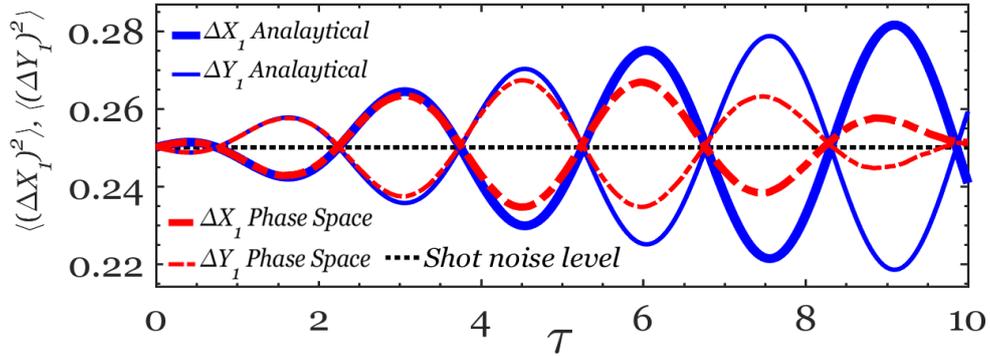

**Figure 2.** Evolution of squeezing for mode $\hat{a}_1$ with $\alpha_1 = \alpha_2 = \alpha_3 = 1$, $\widetilde{\omega}_1 = \widetilde{\omega}_2 = \widetilde{\omega}_3 = 1$, $\tilde{g} = 0.01$ and $\tilde{\kappa} = 0.1$.

In Fig. 3, the results are obtained by initialising the system with a single coherent mode ($\alpha_1 = 1$) and vacuum states for the rest ($\alpha_2 = \alpha_3 = 0$). Additional input parameters are maintained in the same manner as in Fig. 2. Figure 3(a) depicts the squeezing generated by the mode initiated by a coherent mode, while Figures 3(b) and 3(c) illustrate the squeezing generated by the other two modes initiated by a vacuum state. Regarding the coherent mode, both methods are nearly in total agreement and sufficiently in-phase for lengthy periods (up to $\tau \approx 7$). After that, discrepancies between both methods start appearing. Nevertheless, modes prepared in vacuum states have shown an excellent agreement for longer evolution distances, as shown in Fig. 3(b) for vacuum mode $\hat{a}_2$. Similar results were obtained for the vacuum mode $\hat{a}_3$. This supports our hypothesis that the perturbative method yields favorable performance for low-input settings.

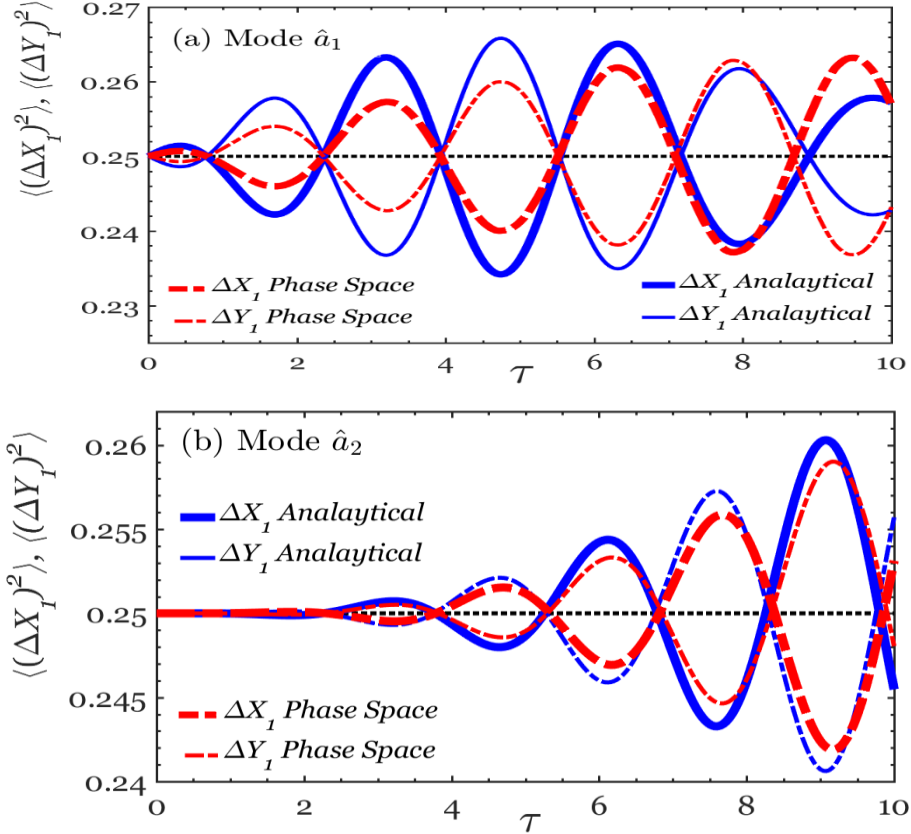

**Figure 3.** Fluctuation of squeezing for (a) coherent-mode $\hat{a}_1$, (b) vacuum-mode $\hat{a}_2$ with $\alpha_1 = 1$, $\alpha_2 = \alpha_3 = 0$, $\widetilde{\omega}_1 = \widetilde{\omega}_2 = \widetilde{\omega}_3 = 1$, $\tilde{g} = 0.01$ and $\tilde{\kappa} = 0.1$.

We now examine the effects of different linear coupling strengths on squeezing. In practice, this parameter can be manipulated by altering the distance between channel waveguides. It is possible to enhance the linear coupling constant by bringing the channel waveguides closer together, and vice versa. Figures 4(a), 4(b), and 4(c) depict the outcomes for scaled linear coupling strength of $\tilde{\kappa} = 0.01$, $\tilde{\kappa} = 0.09$ and $\tilde{\kappa} = 0.5$, respectively. To generate these results, all modes are prepared in coherent states with $\alpha_1 = \alpha_2 = \alpha_3 = 1$ and a common frequency $\widetilde{\omega}_1 = \widetilde{\omega}_2 = \widetilde{\omega}_3 = 1$. In addition, the nonlinear coupling constant is set to $\tilde{g} = 0.01$. Here, we present the outcome for the first mode only. At $\tilde{\kappa} = 0.01$ (Fig. 4(a)), both methods are totally in-phase and completely agree with one another; however, at $\tilde{\kappa} = 0.09$, the agreement in maximal amplitude is only maintained up to $\tau \sim 4$ (Fig. 4(b)). When $\tilde{\kappa} = 0.5$ (Fig. 4(c)), the results obtained from the phase space method and the analytical perturbative method are completely different; the phase space method generates collapses-revivals behaviour with a small amplitude, whereas the analytical perturbative method predicts a very high squeezing amplitude. For $\tilde{\kappa} = 0.5$, the highest level of agreement is only reached during the earliest phases of evolution with a brief interaction period. In addition, the agreement between the two approaches decreases as the linear coupling coefficients increase.

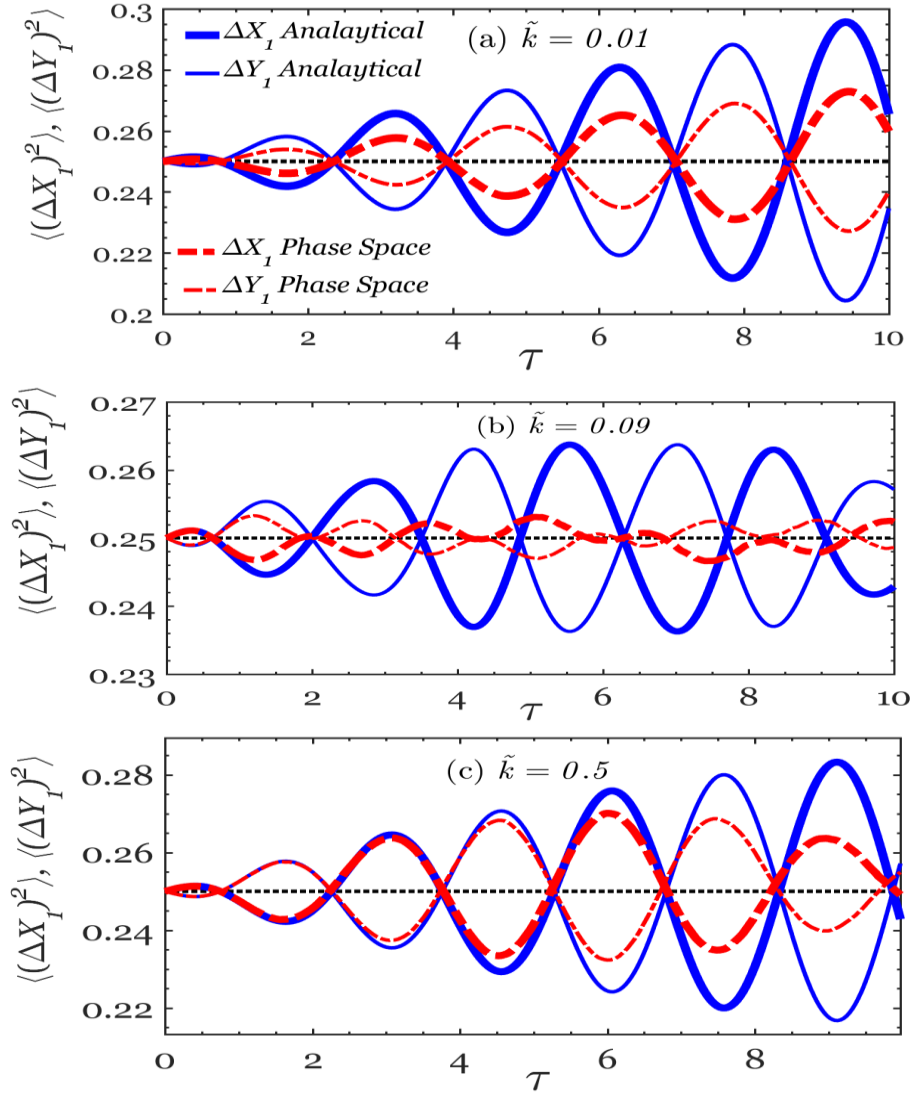

**Figure 4.** Squeezing in mode $\hat{a}_1$ for (a) $\tilde{\kappa} = 0.01$, (b) $\tilde{\kappa} = 0.09$ and (c) $\tilde{\kappa} = 0.5$ with $\alpha_1 = \alpha_2 = \alpha_3 = 1$, $\tilde{\omega}_1 = \tilde{\omega}_2 = \tilde{\omega}_3 = 1$ and $\tilde{g} = 0.01$.

Figure 5 shows squeezing induced by linear coupling constant ($\tilde{k}$) values, as those in Fig. 4. Here, however, the second and third modes are initially created in vacuum states, and we just consider the squeezing caused by the coherent mode. At $\tilde{\kappa} = 0.01$ (Fig. 5(a)), the observed results are comparable to those in Fig. 4(a). This similarity may be attributable to the negligible levels of linear coupling strength. At $\tilde{\kappa} = 0.09$ (Fig. 5(b)), both approaches generate comparable stable oscillation patterns, however, at $\tilde{\kappa} = 0.5$ (Fig. 5(c)), the oscillation of squeezing is no longer continuous, and a pattern of mild collapses and revivals is observed. In addition, a large value of $\tilde{\kappa}$ produces a squeezing with substantial oscillatory quadrature fluctuations, as shown in Fig. 4(c) and 5(c).

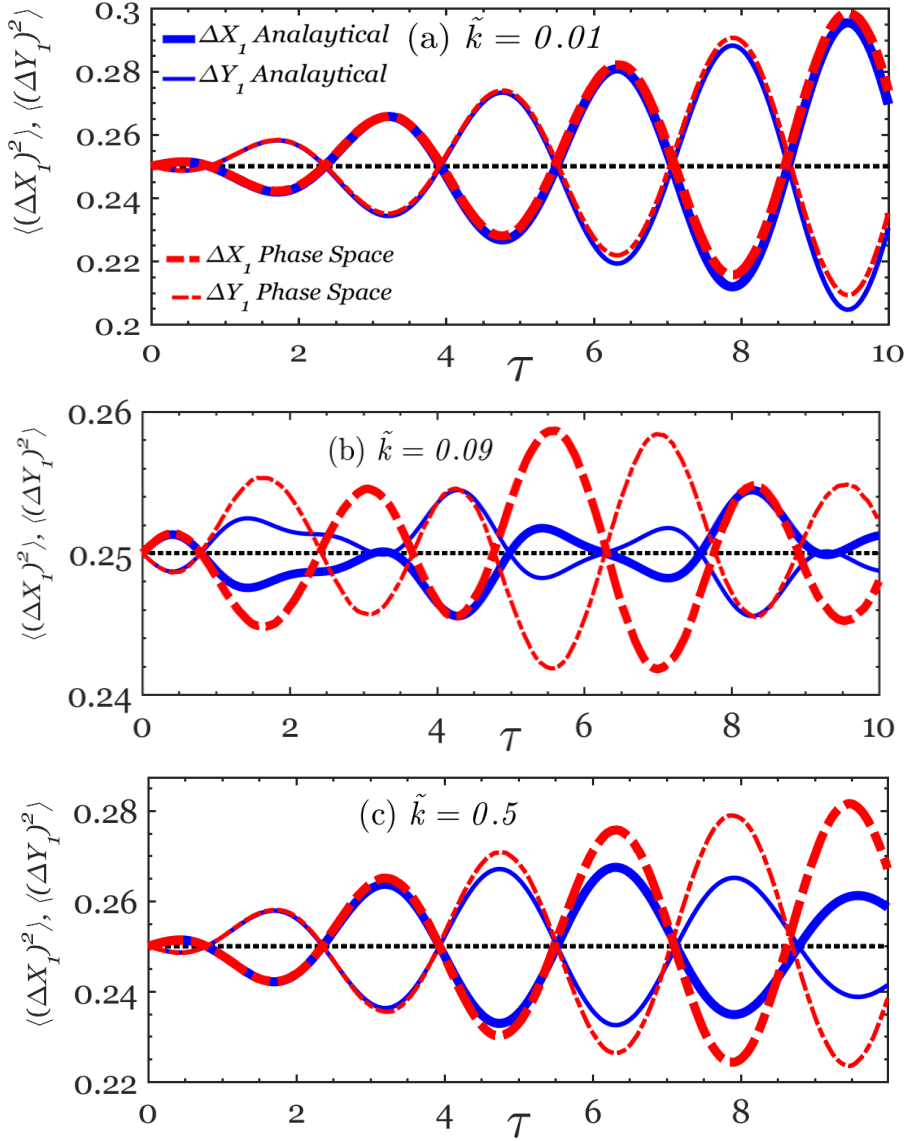

**Figure 5** Single-mode squeezing in coherent-mode $\hat{a}_1$ for (a) $\tilde{\kappa} = 0.01$, (b) $\tilde{\kappa} = 0.09$ and (c) $\tilde{\kappa} = 0.5$ with $\alpha_1 = 1$, $\alpha_2 = \alpha_3 = 0$, $\tilde{\omega}_1 = \tilde{\omega}_2 = \tilde{\omega}_3 = 1$ and $\tilde{g} = 0.01$.

Next, we investigate the influence of nonlinear coupling on the generated squeezed light in the current system. Figures 6 and 7, respectively, depict this for the symmetrical and asymmetrical initializations. Figure 6(a) and (b) illustrate the squeezing produced at two distinct values of coupling coefficient, viz. $\tilde{g} = 0.05$ (Fig. 6(a)) and $\tilde{g} = 0.08$ (Fig. 6(b)). As shown in Fig. 6, both the values of $\tilde{g}$ yield two patterns with similar characteristics. Yet, the maximal squeezing amplitude appears to be greater at $\tilde{g} = 0.08$ than at $\tilde{g} = 0.05$. Hence, the squeezed signal becomes increasingly significant as the nonlinear coupling increases. Beginning at a specific time ($\tau \approx 8$), the squeezing signal disappears in the phase space.

Figure 7 examines the effect of nonlinear coupling in the situation of asymmetrical initialization; the first waveguide is prepared in a coherent state while the other two waveguides are prepared in vacuum states. Like the values used in Fig. 6, squeezing is examined at $\tilde{g} = 0.05$ and $\tilde{g} = 0.08$ (Fig. 7(a) and (b) correspondingly). As anticipated, both values of the nonlinear coefficient result in the same degree of squeezing. In this case, a strong signal of squeezing is also noticed when $g = 0.08$ rather than $g = 0.05$. In conclusion, the intensity of the squeezing grows as the nonlinear coupling rises.

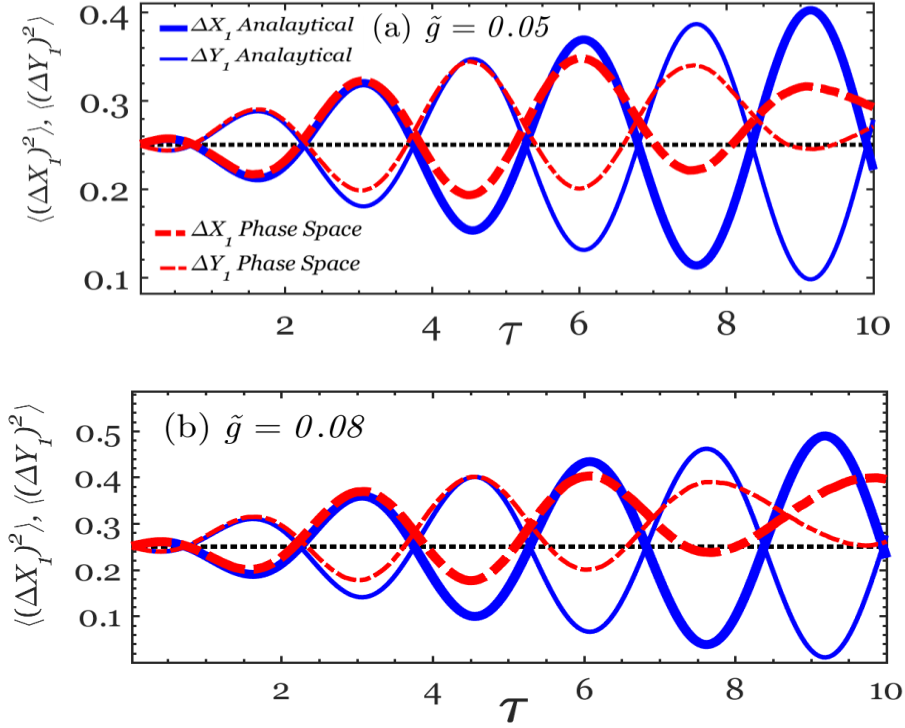

**Figure 6.** Variation of squeezing in mode $\hat{a}_1$ for (a) $\tilde{g} = 0.05$ and (b) $\tilde{g} = 0.08$ with $\alpha_1 = \alpha_2 = \alpha_3 = 1$, $\widetilde{\omega}_1 = \widetilde{\omega}_2 = \widetilde{\omega}_3 = 1$ and $\tilde{\kappa} = 0.1$.

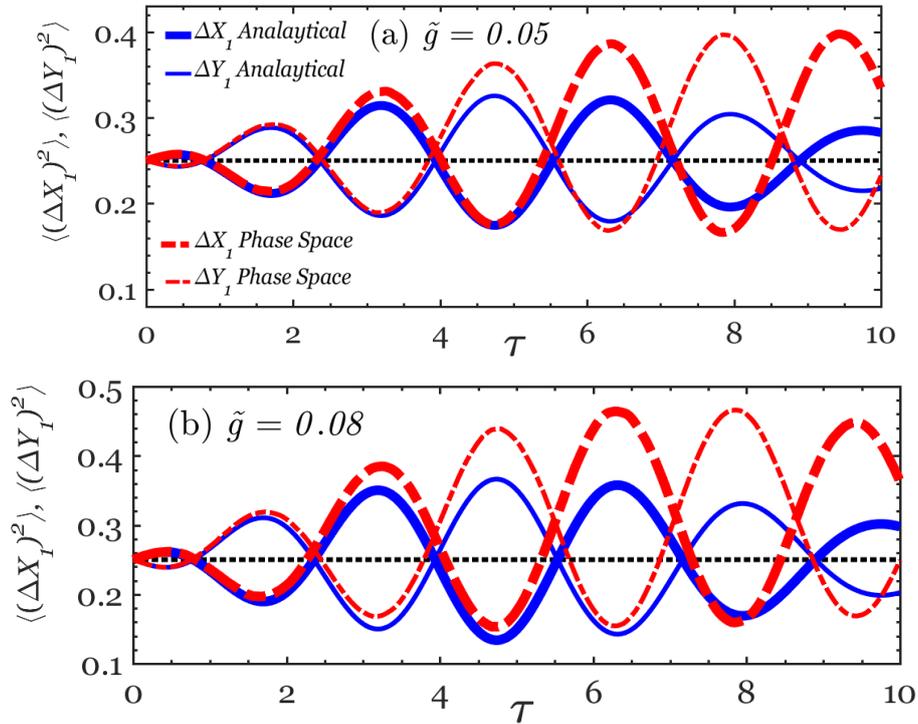

**Figure 7.** Variation of quadrature variances in coherent-mode $\hat{a}_1$ for (a) $\tilde{g} = 0.05$ and (b) $\tilde{g} = 0.08$ with $\alpha_1 = 1$, $\alpha_2 = \alpha_3 = 0$, $\tilde{\omega}_1 = \tilde{\omega}_2 = \tilde{\omega}_3 = 1$ and $\tilde{\kappa} = 0.1$.

Finally, we will present our observation when from the contra-directional propagation. In this case, the second mode is set to be propagating in the opposite direction to the other modes. In Figure 8, the system is initialized symmetrically with all modes having the same amplitude of $\alpha = 1$, and a common wavenumber, $\tilde{k} = 1$. The nonlinear coupling constant is set to be $\tilde{G} = 0.01$, while the linear coupling constant is varied as $\widetilde{K} = 0.1$ (figure 8(a)), $\widetilde{K} = 0.3$ (figure 8(b)) and $\widetilde{K} = 0.5$ (figure 8(c)). As in the codirectional setup, the phase space method agrees very well with the analytical method at early evolution distances. The agreement lasts longer at lower values of the linear couplings (figure 8(a)). Another noticeable difference is that for the phase space method, squeezing with steady oscillation is observed at $\widetilde{K} = 0.1$. When $\widetilde{K}$ increases, the signal becomes lower in amplitude at certain wave periods, perhaps due to some kind of destructive interference. Oppositely, the analytical perturbative method remains less sensitive to the contra-directional arrangement with steady fluctuation of squeezing at $\widehat{K} = 0.1$ and $\widehat{K} = 0.3$, but showing mild collapses and revivals type of pattern at $\widehat{K} = 0.5$.

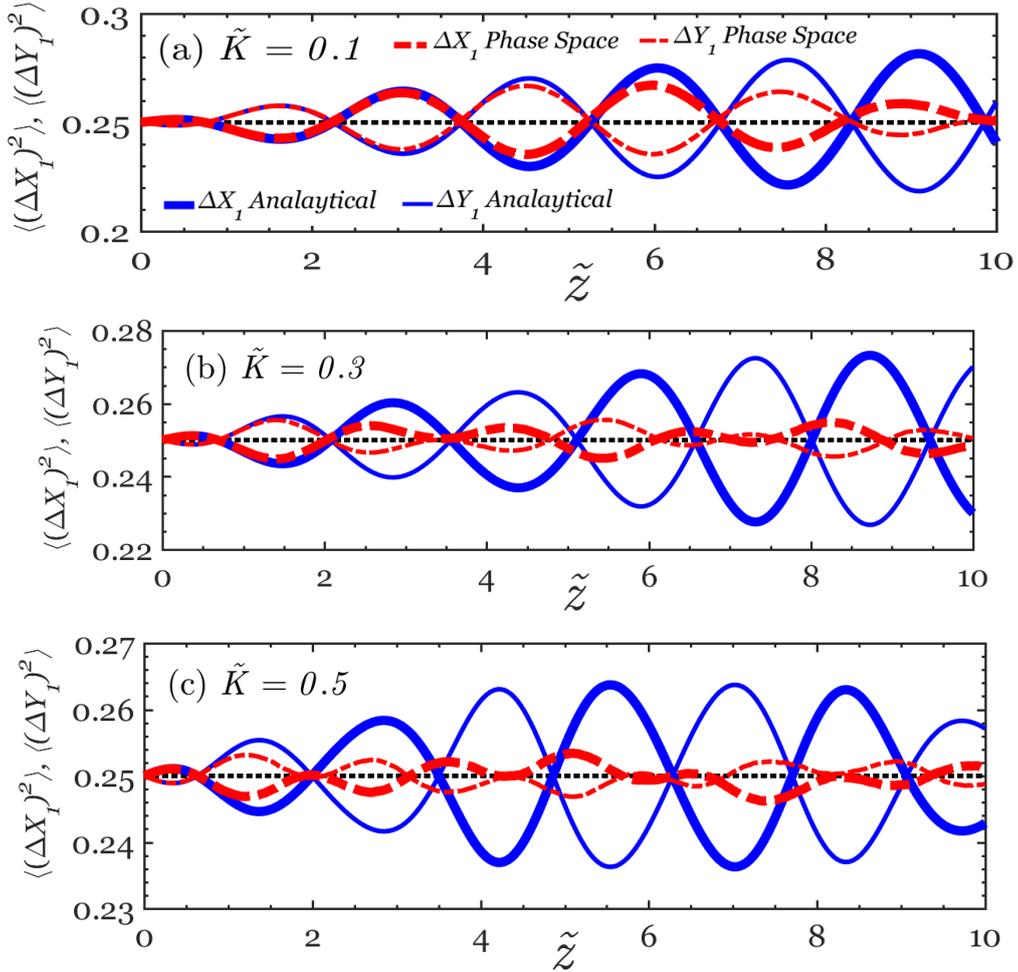

## 4. Conclusion

The squeezed state of light generated in a three-waveguide nonlinear coupler with SH generation has been studied using both the phase space and the analytical perturbative method.

The analytical perturbative method belongs to the Heisenberg picture in which the field operators change over time, but the state vector remains static. The phase space method is based on the Schrödinger picture where the operators remain fixed, but the state vector evolves with time. Rather than depending on a single theoretical prediction, both methods have been used to investigate the squeezed states of light propagating in the current system to ensure accurate findings and to provide insights into the strengths and weaknesses of each method. The effect of key design parameters on the generated squeezed states has been examined and the optimal parameters for the best possible squeezing have been identified. At low values of linear coupling, both methods are in-phase and completely agree with one another. When the linear coupling increases, results obtained exploiting the two approaches differ; the phase space method generates collapses-revivals behavior with a small amplitude, whereas the analytical method predicts a constant increase in the field amplitude. The agreement between the two approaches decreases as the linear coupling coefficients are raised in both codirectional and contra-directional systems. In addition, a stronger linear coupling produces a squeezing with substantial oscillatory quadrature fluctuations. As the nonlinear coupling increases, the degree of squeezing is significantly enhanced. However, this comes at the expense of squeezing range over time. Generally, the perturbative method yields favourable performance for low-input parameters and short periods of evolution. In fact, at certain combinations of design parameters, the analytical method exhibits a constant increase in field amplitude over time which does not appear reasonable. While we think that the positive P phase space method yields more accurate results, the analytical method could be used for short evolution distances and as a verification of the accuracy of the overall results. This opens a new avenue of utilizing NLC in nonclassical light generation in the new era of quantum-based technology. Moreover, the proposed system can serve as the basis for dense optical networks with high-quality data transport.

**Appendix A**

The time-dependent coefficients $\{A_k(t)\}_1^{10}$, $\{B_k(t)\}_1^{10}$, $\{C_k(t)\}_1^{10}$, $\{D_k(t)\}_1^{6}$, $\{E_k(t)\}_1^{6}$ and $\{F_k(t)\}_1^{6}$ appears in the solutions given in (25)-(30) of the perturbative approach, are evaluated precisely by substituting the previous solutions (25)-(30) back into the evolution equations (18)-(23) and equating similar terms on both sides. The sets of coupled equations representing the evolution of these coefficients are obtained as following.

The evolution of the $\{A_k\}_1^{10}$ for fundamental mode $\hat{a}_1$

$$\frac{dA_1}{dt} = -i\omega_1 A_1 - ik\,B_1 - ik\,C_1 \quad \text{(A1)} \qquad \frac{dA_6}{dt} = -i\omega_1 A_6 - ik\,B_7 - ik\,C_7 \quad \text{(A6)}$$

$$\frac{dA_2}{dt} = -i\omega_1 A_2 - ik\,B_2 - ik\,C_2 \quad \text{(A2)} \qquad \frac{dA_7}{dt} = -i\omega_1 A_7 + g\,A_2^*\,D_1 \quad \text{(A7)}$$

$$\frac{dA_3}{dt} = -i\omega_1 A_3 - ik\,B_3 - ik\,C_3 \quad \text{(A3)} \qquad \frac{dA_8}{dt} = -i\omega_1 A_8 + g\,A_3^*\,D_1 \quad \text{(A8)}$$

$$\frac{dA_4}{dt} = -i\omega_1 A_4 + g A_1^* D_1 - ik B_5 - ik C_6 \quad \text{(A4)} \qquad \frac{dA_9}{dt} = -i\omega_1 A_9 + g A_4^* D_1 \quad \text{(A9)}$$

$$\frac{dA_5}{dt} = -i\omega_1 A_5 - ik B_6 - ik C_5 \quad \text{(A5)} \qquad \frac{dA_{10}}{dt} = -i\omega_1 A_{10} + g A_1^* D_2 \quad \text{(A10)}$$

The evolution of the $\{B_k\}_1^{10}$ for fundamental mode $\hat{a}_2$

$$\frac{dB_1}{dt} = -i\omega_2 B_1 - ik A_1 - ik C_1 \quad \text{(A11)} \qquad \frac{dB_6}{dt} = -i\omega_2 B_6 + g B_2^* E_1 - ik A_5 - ik C_5 \quad \text{(A16)}$$

$$\frac{dB_2}{dt} = -i\omega_2 B_2 - ik A_2 - ik C_2 \quad \text{(A12)} \qquad \frac{dB_7}{dt} = -i\omega_2 B_7 - ik A_6 - ik C_7 \quad \text{(A17)}$$

$$\frac{dB_3}{dt} = -i\omega_2 B_3 - ik A_3 - ik C_3 \quad \text{(A13)} \qquad \frac{dB_8}{dt} = -i\omega_2 B_8 + g B_3^* E_1 \quad \text{(A18)}$$

$$\frac{dB_4}{dt} = -i\omega_2 B_4 + g B_1^* E_1 \quad \text{(A14)} \qquad \frac{dB_9}{dt} = -i\omega_2 B_9 + g B_6^* E_1 \quad \text{(A19)}$$

$$\frac{dB_5}{dt} = -i\omega_2 B_5 - ik A_4 - ik C_6 \quad \text{(A15)} \qquad \frac{dB_{10}}{dt} = -i\omega_2 B_{10} + g B_2^* E_2 \quad \text{(A20)}$$

The evolution of the $\{C_k\}_1^{10}$ for fundamental mode $\hat{a}_3$

$$\frac{dC_1}{dt} = -i\omega_3 C_1 - ik B_1 - ik A_1 \quad \text{(A21)} \qquad \frac{dC_6}{dt} = -i\omega_3 C_6 - ik B_5 - ik A_4 \quad \text{(A26)}$$

$$\frac{dC_2}{dt} = -i\omega_3 C_2 - ik B_2 - ik A_2 \quad \text{(A22)} \qquad \frac{dC_7}{dt} = -i\omega_3 C_7 + g C_3^* F_1 - ik B_7 - ik A_6 \quad \text{(A27)}$$

$$\frac{dC_3}{dt} = -i\omega_3 C_3 - ik B_3 - ik A_3 \quad \text{(A23)} \qquad \frac{dC_8}{dt} = -i\omega_3 C_8 + g C_7^* F_1 \quad \text{(A28)}$$

$$\frac{dC_4}{dt} = -i\omega_3 C_4 + g C_1^* F_1 \quad \text{(A24)} \qquad \frac{dC_9}{dt} = -i\omega_3 C_9 + g C_2^* F_1 \quad \text{(A29)}$$

$$\frac{dC_5}{dt} = -i\omega_3 C_5 - ik B_6 - ik A_5 \quad \text{(A25)} \qquad \frac{dC_{10}}{dt} = -i\omega_3 C_{10} + g C_3^* F_4 \quad \text{(A30)}$$

The evolution of the $\{D_k(t)\}_1^6$ for the first harmonic mode $\hat{b}_1$

$$\frac{dD_1}{dt} = -i(2\omega_1)D_1 \quad \text{(A31)} \qquad \frac{dD_4}{dt} = -i(2\omega_1)D_4 - g A_1 A_3 \quad \text{(A34)}$$

$$\frac{dD_2}{dt} = -i(2\omega_1)D_2 - \frac{g}{2} A_1^2 \quad \text{(A32)} \qquad \frac{dD_5}{dt} = -i(2\omega_1)D_5 - \frac{g}{2} A_4 A_1 \quad \text{(A35)}$$

$$\frac{dD_3}{dt} = -i(2\omega_1)D_3 - g A_1 A_2 \quad \text{(A33)} \qquad \frac{dD_6}{dt} = -i(2\omega_1)D_6 - \frac{g}{2} A_1 A_4 \quad \text{(A36)}$$

The evolution of the $\{E_k(t)\}_1^6$ for the SH mode $\hat{b}_2$

$$\frac{d\,E_1}{dt} = -i(2\omega_2)E_1 \qquad \text{(A37)}$$

$$\frac{d\,E_4}{dt} = -i(2\omega_2)E_4 - g\,B_2 B_3 \qquad \text{(A40)}$$

$$\frac{d\,E_2}{dt} = -i(2\omega_2)E_2 - \frac{g}{2}B_2^{\ 2} \qquad \text{(A38)}$$

$$\frac{d\,E_5}{dt} = -i(2\omega_2)E_5 - \frac{g}{2}B_6 B_2 \qquad \text{(A41)}$$

$$\frac{d\,E_3}{dt} = -i(2\omega_2)E_3 - g\,B_1 B_2 \qquad \text{(A39)}$$

$$\frac{d\,E_6}{dt} = -i(2\omega_2)E_6 - \frac{g}{2}B_2 B_6 \qquad \text{(A42)}$$

The evolution of the $\{F_k(t)\}_1^6$ for the third harmonic mode $\hat{b}_3$

$$\frac{d\,F_1}{dt} = -i(2\omega_3)F_1 \qquad \text{(A43)}$$

$$\frac{d\,F_4}{dt} = -i(2\omega_3)F_4 - \frac{g}{2}C_3^{\ 2} \qquad \text{(A46)}$$

$$\frac{d\,F_2}{dt} = -i(2\omega_3)F_2 - g\,C_2 C_3 \qquad \text{(A44)}$$

$$\frac{d\,F_5}{dt} = -i(2\omega_3)F_5 - \frac{g}{2}C_7 C_3 \qquad \text{(A47)}$$

$$\frac{d\,F_3}{dt} = -i(2\omega_3)F_3 - g\,C_1 C_3 \qquad \text{(A45)}$$

$$\frac{d\,F_6}{dt} = -i(2\omega_3)F_6 - \frac{g}{2}C_3 C_7 \qquad \text{(A48)}$$

## Declarations

**Availability of data and materials**

A MATLAB code was developed to generate these theoretical results. The code is available from the corresponding author upon reasonable request.

**Competing interests**

The authors declare no competing interests.

**Funding**

R Julius gratefully recognizes the Malaysian Ministry of Higher Education (MOHE) national grant FRGS/1/2021/STG07/UITM/02/4.

**Authors' contributions**

The main idea, mathematical derivation, numerical simulation, and results were contributed by M.S.M.H, A.M.A.I., and R.J respectively. while A.M.A.I., P.K.C. and H. E. looked over the entire study and the manuscript preparation.

**Acknowledgments**

Not applicable